\documentstyle[twoside,epsf]{kaeog}

\begin{document}

\author{Marek Gutowski\\
	Institute of Physics,
	Polish Academy of Sciences\\
	02--668 Warszawa, Al. Lotnik\'ow 32/46,
	Poland\\
	e-mail: gutow@ifpan.edu.pl
}

\title{L\'evy flights as an underlying mechanism for global
optimization algorithms}


\maketitle


\begin{abstract}

In this paper we propose and advocate the use of the so called {\em
L\'evy flights\/} as a driving mechanism for a class of stochastic
optimization computations. This proposal, for some reasons overlooked
until now, is -- in author's opinion -- very appropriate to satisfy the
need for algorithm, which is capable of generating trial steps of very
different length in the search space.  The required balance between
short and long steps can be easily and fully controlled.  A~simple
example of approximated L\'evy distribution, implemented in {\sf
FORTRAN 77\/}, is given.  We also discuss the physical grounds of
presented methods.

\end{abstract}



\section{Optimization algorithms and physics}

More and more global optimization tasks are completed today using
algorithms originating from mimicking the way the Nature solves them. 
We have two branches of science, which describe the world around us,
namely biology (for living things) and physics (for the rest of it). 
The problem of global optimization (either minimization or
maximization, with or without constraints) of the objective function of
many variables still remains a~challenge for practitioners.  There is
no single, universal and deterministic algorithm capable of solving all
kinds of optimization problems: those involving smooth as well as
non-smooth objective functions, mixed integer-real-boolean valued
unknowns, etc.

Classical mathematical analysis was long the only tool for finding
extrema of functions of many variables. Unfortunately it is of very
limited use in many practical applications, especially when the
objective function is not differentiable at least once. On the other
hand, we are ready to accept the solutions, which are not perfect in
the mathematical sense (often called "$\varepsilon$-optimal"), but are
sufficiently close to them.

To overcome such difficulties, researchers in various fields of science
and engineering turned to stochastic algorithms.  There are several
kinds of arguments for doing so. The first, and certainly not the most
important one, is increasing availability of the computing power. We
are able to examine, usually in a~fraction of a~second, many trial
solutions of the optimization problem under study.  Loosely speaking,
this is the base of a~rich family of Monte-Carlo-type optimization
algorithms.  Trying to mimic Nature's actions is another justification
for rich variety of optimization algorithms, just because Nature seems
very successful.  Let's put aside algorithms of genetic type, grown on
biological grounds, and concentrate instead on those, which are based
on behavior of purely physical systems.  Various physical phenomena
were taken into account, mostly from classical mechanics of a~single
particle (deterministic algorithms of "gradient type") and
thermodynamics (simulated annealing) as models for optimization
procedures.

In this paper we propose the diffusion processes and quantum tunneling
as a base for a class of stochastic optimization routines.

\section{Diffusive processes as a model for optimization procedure}

Consider the simplest version of familiar Monte-Carlo optimization
procedure.  Its ope\-ra\-tion may be summarized as a~random walk in the
search space (bounded or not) and sampling the values of objective
function in visited points.  The trajectory of such a~random walker is
very similar to the trajectory of physical particle subjected to
Brownian motion.  If we consider many random walkers at the same time
(multiple start point Monte-Carlo optimization algorithm), then
emerging set of trajectories resembles closely the diffusion process.
In the derivation of the law of Brownian motion one assumes that the
lengths of individual "jumps" are not equal to each other but are
distributed normally, as a result of a huge number (estimated as
$10^{5}$---$10^{6}$) of independent "kicks" from surrounding molecules. 
This is practical manifestation of the law of large numbers, known also
as the Central Limit Theorem.  Under those assumptions it may be shown
(Einstein, Smoluchowski) that the average distance, $R$, of a random
walker from starting point is a~function of time, $t$, and can be
expressed as
\begin{equation}\label{Brown}
	\langle R^{2}(t)\rangle={\cal D} t^{\nu},\quad {\rm with}\ \nu=1
\end{equation}
where ${\cal D}$ is the diffusion constant.

The formula (\ref{Brown}) was later confirmed experimentally for small
particles suspended in liquids. This, in turn, made possible to
estimate the value of very important physical constant, the Avogadro's
number, $N_{A}$, thus finally confirming the atomic structure of
matter.  It is worth noting that the early value of $N_{A}$, obtained
this way, differs less than $1$\% from the one known today, almost
$100$ years later.

Extensive investigations of diffusion processes revealed, that at least
some of them must be governed by other mechanisms, different from
familiar Brownian motion.  They were classified as {\em subdiffusion\/}
($\nu < 1$, see Eq. \ref{Brown}) and {\em enhanced diffusion\/} ($\nu >
1$).  It is often said that the enhanced diffusion is governed by {\em
L\'evy flights\/}, which will be explained below.

\bigskip\noindent
{\em L\'evy flights}
\medskip

Paul L\'evy (1886---1971), the French mathematician, considered in
thirties (XX century) the following problem \cite{PhysW}:
\begin{quote}
  {\sl What, if any, should be the probability density of $N$ independent,
identically distributed random variables ($iid$) $X_{1}$, $X_{2}$, \ldots,
$X_{N}$ to satisfy the requirement that the probability density of
their sum $X_{1}+X_{2}+\ldots +X_{N}$ has the same functional form?}
\end{quote}
Today we could say, that L\'evy tried to find a class of self-similar
objects, known as {\em fractals\/} since Benoit Mandelbrot had invented
them, much later, in 1968.

Well known answer to L\'evy's problem was based on famous {\em Central
Limit Theorem\/}, which, in most widely known version, states that the
sum of $iid$ random variables has normal probability density.  We can
even drop the requirement of identical distributions (but {\em not\/}
the independence!) and still have the same result. There is a catch,
however: the individual distributions have to be {\em narrow\/}, i.e.
their first and second moment (Lindberg), and in Lyapunov version also
the third moment, have to be finite\footnote{In Lyapunov version we
need even stronger condition, which may be, not quite precisely,
expressed as {\em no random variable dominates others in the sum\/}. 
Finiteness of third moments is necessary but not enough for that.}.

Taking into considerations also other, non-gaussian distributions,
L\'evy obtained the following condition for the Fourier transform of
probability density of the sum of $N$ $iid$ random variables:
\begin{equation}\label{orig}
	\tilde{p}_{N}(k) \sim \exp \left( -N|k|^{\beta} \right)
\end{equation}
where the normalization constant was dropped, and $0 < \beta < 2$.

Going back to the searched distribution, not its Fourier transform, is
not trivial, and the analytical form of the result is known only for
few special cases. Generally it may be expressed as \cite{PRL}
\begin{equation}\label{calka}
	L(x) = \frac{1}{\pi} \int_{0}^{\infty} \exp \left( -\gamma
	q^{\beta} \right) \cos qx \ dq
\end{equation}
and is known as {\em symmetrical L\'evy stable distribution of index\/}
$\beta$ ($0 < \beta \le 2$) {\em and scale factor\/} $\gamma$ ($\gamma
> 0$).  For simplicity one usually sets $\gamma=1$.

\medskip
The special cases mentioned earlier are:
\begin{itemize}
\item 
Cauchy distribution (among physicists known also as Lorentzian shape):
\begin{equation}
	p_{N}(x) = \frac{1}{\pi N}\
	\frac{1}{1+\left(\frac{x}{N}\right)^{2}} = \frac{1}{N}\
	p_{1}\left(\frac{x}{N}\right)\quad\quad {\rm for\ }\beta=1,\
	{\rm and}
\end{equation}
\item
Gauss normal distribution, when $\beta=2$.
\end{itemize}

The integral (\ref{calka}) can be written in a form of (truncated)
power series \cite{Berg}
\begin{equation}\label{szereg}
	L(x) = -\frac{1}{\pi} \sum_{k=1}^{m}
	\frac{\left(-1\right)^{k}}{k!} \frac{\Gamma\left(\beta
	k+1\right)}{x^{\beta k+1}} \sin
	\left(\frac{k\pi\beta}{2}\right) + R_{m}(x)
\end{equation}
with $R_{m}(x)$ of order $x^{-\beta\left(m+1\right)-1}$ and the leading
term is proportional to $x^{-1-\beta}$.  Looking at the above series
and original L\'evy's result (\ref{orig}), one can see that the
searched probability density should behave as
\begin{equation}\label{approx}
	L(x) \sim |x|^{-1-\beta}\quad {\rm as\ } |x| \rightarrow \infty
\end{equation}
Now we understand, why the index $\beta$ {\em must\/} belong to the
interval $\left.\right]0,2\left.\right]$: for $\beta \le 0$ integral
(\ref{calka}) does not exist (is unbounded), while for $\beta \ge 2$
ordinary {\em Central Limit Theorem\/} holds.  It is also clear, that
there are some L\'evy distributions, those with index $0 < \beta < 1$,
for which even the first moment, i.e. expectation value, does not exist
(second moment, i.e. the variance, is always infinite).  This poses a
serious problem for physicists, since the ordinary procedure of
repeated measurements makes no sense in such cases, and if used
nevertheless -- leads to strange, confusing and incorrect results.

\section{Diffusive processes, continued}
Allowing the random walker to make steps of length $l$
distributed\footnote{To be precise: we should write $l/l_{0}$ here,
instead of just $l$, where $l_{0}$ denotes unit length, in order to
operate with dimensionless quantities only.  The reason for not doing
so is following: we don't want to create the impression that any
particular length scale is better than others; indeed any unit ranging
from femtometers to astronomical unit is equally good.  That is why the
behavior described by power law is often called 'scale free' -- no
length unit is preferred, except for practical reasons.} as
\begin{equation}\label{easy}
	P(l)=\frac{C}{(1+l)^{1+\beta}}
\end{equation}
with appropriate normalization constant $C$, one can show that the
interesting quantity $\langle R^{2}(t)\rangle$ follows the law
\begin{equation}
	\langle R^{2}(t)\rangle \sim \left\{
\begin{array}{ll}
t^{2}                    & 0<\beta<1 \nonumber \\
t^{2}/\ln t              & \beta=1   \nonumber \\
t^{3-\beta}              & 1<\beta<2 \nonumber \\
t \ln t                  & \beta=2   \nonumber \\
t                        & \beta>2
\end{array}
	\right.
\end{equation}
assuming that $l=vt$, and $v=const>0$ during every jump.

The distribution (\ref{easy}) approximates very well the L\'evy
distribution for large arguments, see relation (\ref{approx}).  Some
authors prefer to use another approximation for L\'evy distribution:
$P(l) \sim\ {l^{-(1+\mu})}$ (for large $l$), inferring then that
$\langle R^{2}(t)\rangle \sim t^{2/\mu}$.  They use $P(l)=const$ for
small $l$.  We prefer our form, since it never produces infinite
densities of probability, while retaining desired asymptotic properties.

\begin{figure}[h]
  \epsfxsize=1.05\hsize
  \epsfysize=0.65\hsize
  \centerline{\epsfbox{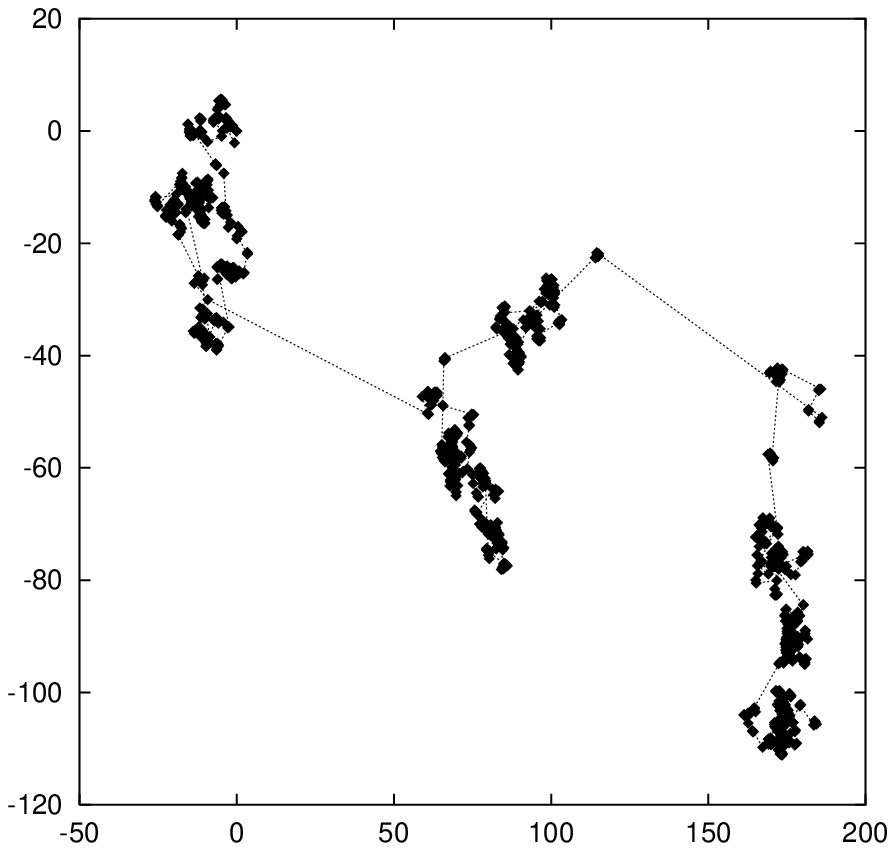}}
\caption{Example of L\'evy flight with index $\beta = 1.99<2$.  Here,
and in subsequent figures, presented are the trajectories consisting
of~$500$ straight sections, obtained with power law distribution of
step sizes approximating L\'evy distribution, see text, and uniform
distribution of directions in plane.  The sequences of numbers produced
by underlying uniform random number generator are identical in all
figures and the random walk always starts from position $(0,0)$.
Clearly visible are characteristic features of trajectories: they are
clusters of clusters of clusters of\ \ldots\ }
\label{L1.99}
\end{figure}

It is interesting, that in such a vigorous movement as the turbulence,
the squared average distance from start point, for any particle
observed in a coordinate system moving with the fluid (Lagrange's
coordinates), may be characterized by $\beta=5/3$.  Other physical
systems described by L\'evy distribution are mentioned in
\cite{PRL,laser}.  Especially, many physical quantities in phase
transition region behave according to the power law. Among them we can
find relaxing sand piles, magnetic systems, etc.  The lengths of flights
made by albatrosses are also distributed according to power law.  Other
examples from everyday life include stock market price fluctuations,
www network connectivity (number of computers connected to the given
node), compacting the granular systems and\ \ldots\ the number of goals
per soccer game.

\bigskip\noindent
{\em Why L\'evy distribution may be useful for stochastic
optimization?\/} 

\begin{figure}[h]
  \epsfxsize=0.85\hsize
  \epsfysize=0.6\hsize
  \centerline{\epsfbox{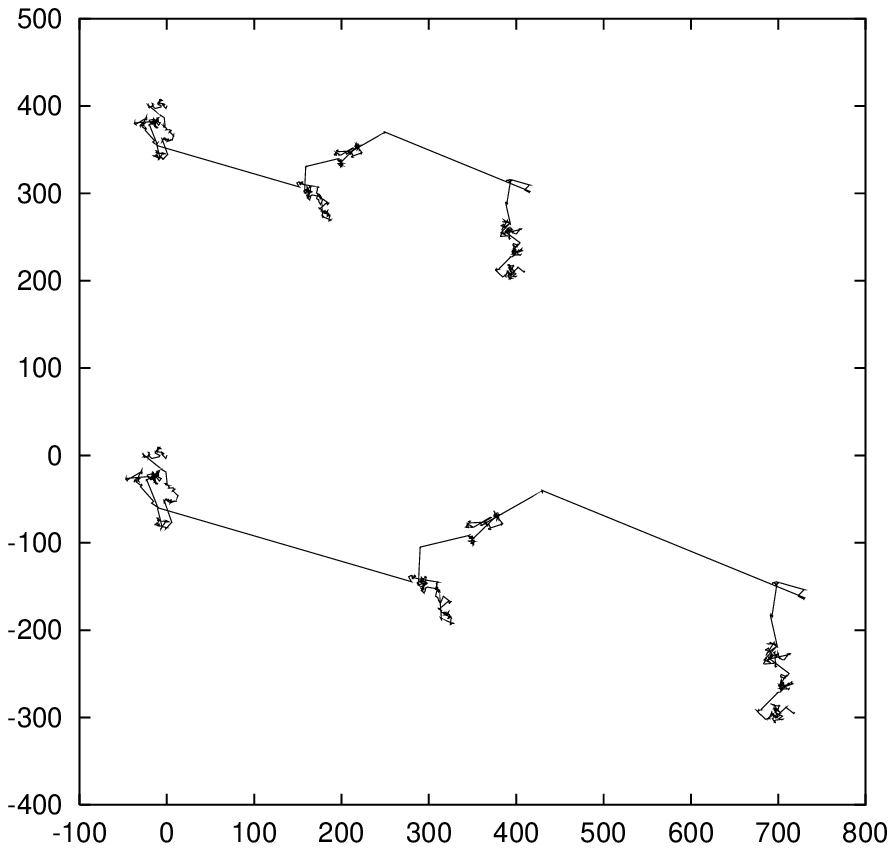}}
\caption{L\'evy flight with index $\beta = 1.67\approx 5/3$ (turbulent
case, upper curve shifted $400$ units up) and $\beta=1.50$ (lower
curve).}
\label{L150+167}
\end{figure}

\begin{figure}[h]
  \epsfxsize=1.05\hsize
  \epsfysize=0.408\hsize
  \centerline{\epsfbox{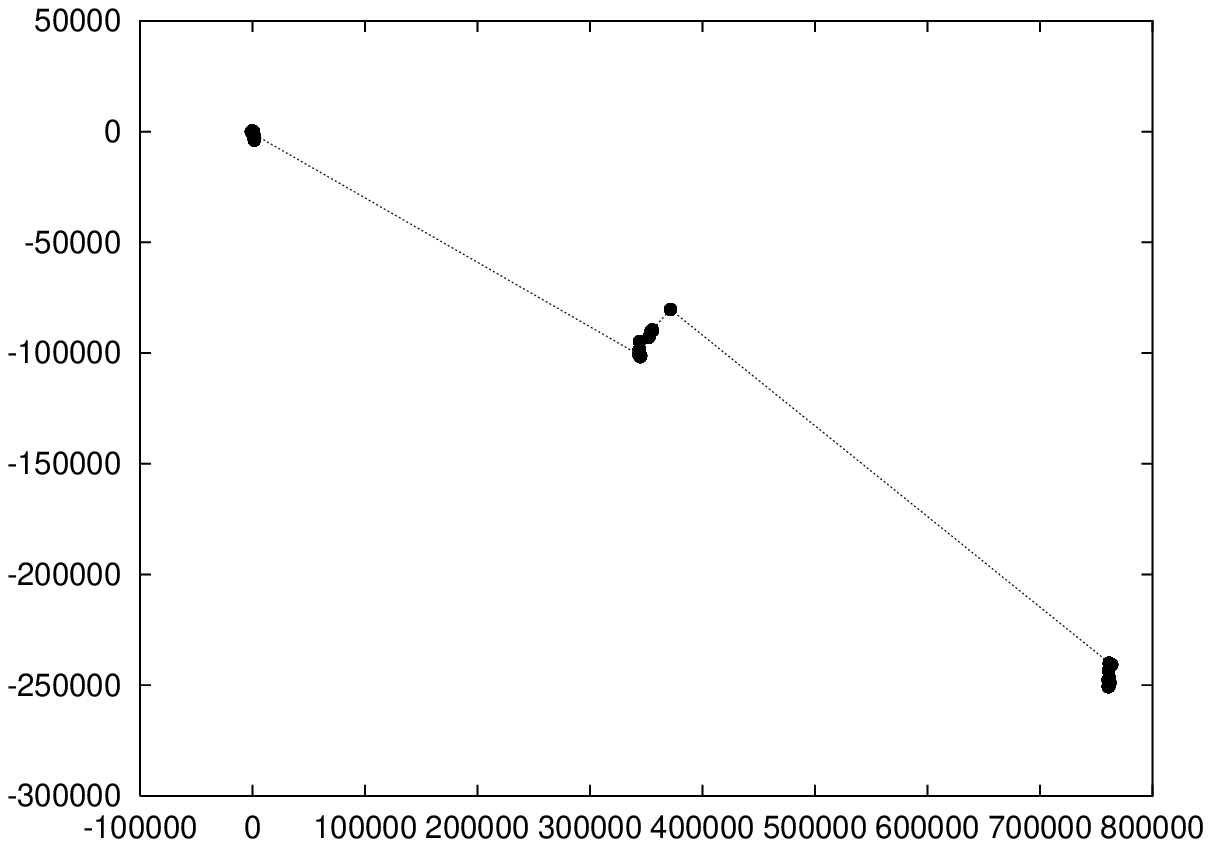}}
\caption{L\'evy flight with index $\beta = 0.67 < 1$.  Note the
dramatic scale change ($\sim 1000\times$) comparing to previous
figures. Contrary to earlier presented cases, the average position does
not exist, so this walker has capability to explore unbounded domains.
Recall that this trajectory also consists of $500$ sections.}
\label{L0.67}
\end{figure}

\medskip
In global stochastic optimization we need two essential ingredients, in
some sense acting against each other. One of them is the routine, which
finds efficiently the local extremum when the search process happens to
be nearby, and the other -- a way to escape from local extremum, since
it may be not the global one.  It is common to observe during
evolutionary optimization the prolonged periods of relatively small
improvements followed by sudden, rapid transitions to another local
extremum.  The process may take long time simply because the random
walkers move and explore the search space too slowly, i.e. they make
too small and hence too cautious steps.  Using step size generated
accordingly to one of L\'evy distributions instead of uniform or
gaussian distribution should therefore be advantageous.  The population
of random walkers {\bf will be always concentrated around recently
found extremum}, as it should in evolutionary algorithms, and in the
same time always few population members {\bf will explore more distant
regions of search space}.  This happens with normally distributed
walkers only {\em very\/} rarely.  The ratio between two classes of
random walkers may be easily controlled, in a smooth way, by
appropriate choice of index $\beta$.  This is illustrated in Figs.
\ref{L1.99}---\ref{L0.67}.

True L\'evy distribution is hard to implement in computer code, but the
approximate form, like the one given by Eq. (\ref{easy}), is easy (see
the next section).  The optimal choice of index $\beta$ may be
problem-dependent, but should not be critical.  Further investigations
(experiments?) are necessary to address this question.  We suppose,
that this choice should be concentrated mainly in the range $\left.
\right]0,1\left. \right]$, as largely unexplored until now.  Values
$\beta > 1$ result in slower spreading of random walkers in search
space, what may very significantly affect their ability to find desired
extremum, when the problem is defined on unbounded domain.  On the
other hand, the case $0<\beta<1$ may be considered as a computer
imitation of the phenomenon known from quantum mechanics, namely {\bf
quantum tunneling}, see Fig.~\ref{L0.67}.  It is interesting to note,
that we have obtained this behavior without even mentioning the quantum
mechanical methods.

One may wonder, whether such an optimization procedure can still be
classified as an evolutionary algorithm.  Our answer is {\em yes\/},
because one can always identify long jumps in search space with {\em
mutations\/}, but remember that some researchers are simply ruling out
mutations from algorithms regarded as evolutionary, reserving them
instead for --- as they call it --- genetic-type routines.

We have to mention here, that similar behavior, i.e. occasional long
jumps, was introduced earlier, rather heuristically, by Galar
\cite{Galar} and Kopciuch under the amusing name of {\em impatience
operator\/} and interpreted there mainly in context of social sciences.

\section{Example \protect{\sf FORTRAN~77} procedure for generating
L\'evy flights}

Here we present the random number generator, implemented in {\sf
FORTRAN 77}, which produces sequences of numbers distributed according
to Eq. (\ref{easy}).  It is intentionally not optimized and works by
inverting the distributive function, which is given by
\begin{equation}
	{\cal D}^{-1}(\xi) = 1/(1-\xi)^{1/\beta} - 1,
\end{equation}
where $\xi$ is uniformly distributed on $[0,1]$.  The above form may be
safely simplified to
\begin{equation}
	{\cal D}^{-1}(\xi) = \xi^{-1/\beta} - 1
\end{equation}
by replacement $1-\xi \leftrightarrow \xi$.

\medskip
\begin{center}
\begin{minipage}[h]{0.7\hsize}
{\sf {\obeylines \obeyspaces
      DOUBLE PRECISION FUNCTION LEVY1 (X, BETA)
      DOUBLE PRECISION X, BETA, R, RANF
      R = RANF(X)
      LEVY1 = 1.D0/R**(1.D0/BETA) - 1.D0
      RETURN
      END
}}
\end{minipage}
\end{center}

\medskip
{\sf RANF} is the name of any available standard, i.e. uniformly
distributed on $\left.\right]0,1\left.\right]$, random number generator
taking {\sf X} (of type {\sf DOUBLE PRECISION}) as a dummy argument.

There is no check, whether $\beta \in [0,2[$.  The subroutine will work
even for $\beta \ge 2$, however one should not expect to obtain
normally distributed random numbers in such case.

\section{Summary and discussion}
In this paper we have described two distinct, but closely related
classes of stochastic optimization algorithms, based on a~single
mathematical model and being the computer counterparts of two different
physical effects: classical diffusion and quantum tunneling.  They
received unified mathematical background and may be distinguished
according to the properties of random walkers, as summarized below:

\medskip
\begin{center}
\begin{tabular}{lll}
{\bf properties}  & {\bf L\'evy index} & {\bf physical effect}\\
\hline\hline
no moment exists & $0 < \beta \le 1$ & quantum tunneling\\
\hline
only first moment exists & $1 < \beta < 2$ & superdiffusion, including turbulence\\
\hline
gaussian distribution    & $\beta=2$       & diffusion (Brownian motion)\\
\hline
unknown with $\sigma^{2} < \infty$    & $\beta > 2$ & subdiffusion\\
                         & or not applicable  & \\
\hline
\end{tabular}
\end{center}

\medskip
The unexpected similarities between classical diffusion processes and
quantum tunneling have their roots probably in properties
(similarities) of the corresponding partial differential equations
describing them.  Both equations, i.e. the diffusion equation (Fick's
law) and Schr\"odinger equation relate first partial derivatives of the
unknown function with respect to time with its second spatial
derivatives.  The important difference is the explicit presence of
imaginary unit, $i$, in Schr\"odinger's equation.  The algorithm we
describe here is able to mimic the properties of both types of
solutions.  It can be easily switched from one type of behavior to
another one by merely changing the value of a~single control parameter,
i.e. L\'evy index.

The question arises, whether the familiar evolutionary algorithms
should be immediately thrown away in favor of L\'evy flights based
ones.  Even, if we stick to the orthodox definition of evolutionary
algorithms as the ones, which accept only small, gradual changes in
position within the searched domain -- then the answer is {\em no\/}.
The main problem with evolutionary algorithms is their poor ability to
escape from unwanted, suboptimal extrema.  Indeed, if evolutionary
random walkers make steps with lengths distributed uniformly on
$[0,l_{max}]$, then they are unable to escape from extremum, which is
wider than $\sim 2l_{max}$.  Using normally distributed numbers as step
lengths -- good in theory -- doesn't help much in practice.  And here
is why: gaussian generators of poor quality {\em never\/} produce
random numbers exceeding few (say $\sim 3$) $\sigma$ in magnitude.  On
the other hand, even perfect normal generators produce {\em very
rarely\/} steps longer than that.  So it is a~pure illusion, that it is
possible to find the global optimum in reasonable time with one of such
algorithms -- even if the appropriate theorem states so.  This may only
happen, when our first approximation to the solution is already quite
good.  Quite different situation occurs, when our goal is to keep track
and adapt to the varying environment, i.e. when objective function
changes smoothly and slowly enough while the optimization procedure is
in progress (for example satellite or missile tracking).  In such
cases, providing the starting point is already well known -- i.e. is
located closer to the global optimum than to any other one -- the
evolutionary algorithm, without any mutations, is indispensable.

\section{Acknowledgment}
This work was done as part of author's statutory activity in Institute
of Physics, Polish Academy of Sciences.


\end{document}